# Towards Robust Hemolysis Modeling with Uncertainty Quantification: A Universal Approach to Address Experimental Variance

Christopher Blum[a,*], Ulrich Steinseifer[a], Michael Neidlin[a]

**Affiliation:**

a) Department of Cardiovascular Engineering, Institute of Applied Medical Engineering, Medical Faculty, RWTH Aachen University, Aachen, Germany

**\*Correspondence:**

Name: Christopher Blum

Address: Institute of Applied Medical Engineering, Pauwelsstraße 20

52074 Aachen, Germany

Email address: blum@ame.rwth-aachen.de

**Acknowledgments:**

Funded by the Deutsche Forschungsgemeinschaft (DFG, German Research Foundation) – project number 467133626. Simulations were performed with computing resources granted by RWTH Aachen University under project rwth1211. The authors thank Prof. Marek Behr and Nico Dirkes for experimental data.

**Conflict of interest:**

All of the authors have nothing to disclose.

**Authors' contributions:**

All authors contributed to the study conception and design. CB developed the numerical model, performed the simulations, gathered, analyzed and discussed the results. MN and US were involved in the analysis and discussion of the results. MN supervised the project. CB wrote the manuscript based on the input of all co-authors. All co-authors read and approved the final version of the manuscript.

# Abstract

**Purpose:**

The purpose of this study is to address the lack of uncertainty quantification in numerical hemolysis models, which are critical for medical device evaluations. Specifically, we aim to incorporate experimental variability into these models using the Markov Chain Monte Carlo (MCMC) method to enhance predictive accuracy and robustness.

**Methods:**

We applied the MCMC method to an experimental hemolysis dataset to derive detailed stochastic distributions for the hemolysis Power Law model parameters C, α, and β. These distributions were then propagated through a reduced order model of the FDA benchmark pump to quantify the experimental uncertainty in hemolysis measurements with respect to the predicted pump hemolysis.

**Results:**

The MCMC analysis revealed multiple local minima in the sum of squared errors, highlighting the non-uniqueness of traditional Power Law model fitting. The MCMC results showed a constant optimal $C = 3.515 \times 10^{-5}$ and log normal distributions of α and β with means of 0.614 and 1.795, respectively. The MCMC model closely matched the mean and variance of experimental data. In comparison, conventional deterministic models are not able to describe experimental variation.

**Conclusion:**

Incorporating Uncertainty quantification through MCMC enhances the robustness and predictive accuracy of hemolysis models. This method allows for better comparison of simulated hemolysis outcomes with in-vivo experiments and can integrate additional datasets, potentially setting a new standard in hemolysis modeling

**Key Terms:**

Markov Chain Monte Carlo, hemolysis modeling, Uncertainty Quantification Hemolysis, in-silico

# 1. Introduction

Hemolysis, the destruction of red blood cells, presents a significant challenge in the development and use of cardiovascular blood-contacting devices. When red blood cells are damaged, hemoglobin is released into the bloodstream, potentially causing serious complications. This phenomenon is particularly relevant in devices such as mechanical heart valves [1], ventricular assist devices (VADs) [2], and other forms of mechanical circulatory support (MCS) [3, 4]. These life-saving devices can exert mechanical forces that rupture red blood cells, leading to hemolysis. Due to supraphysiological shear rates especially in MCS devices this is a major concern as high levels of hemolysis are associated with increased mortality [5], thrombosis and bleeding events [2, 6]. Therefore, mitigating hemolysis is crucial for improving the safety and efficacy of these devices.

To address this issue, researchers employ in-silico numerical hemolysis models to optimize MCS designs, aiming to reduce adverse events. In-silico models are cost-effective alternatives to in-vitro and in-vivo studies, saving both time and resources, that are applied across various disciplines [7–9]. Recognizing the importance of in-silico models, the FDA has recently issued guidelines supporting the use of computational models for device approvals, emphasizing the need for reliable and accurate computational models [10]. Central to these guidelines are verification, validation, and uncertainty quantification (VV-UQ), which ensure the reliability of the computational model and account for real-world data variations.

Given the emphasis on uncertainty quantification (UQ), it is essential to examine the current computational gold standard for predicting hemolysis: the Power Law method introduced by Giersiepen et al. [11]. This method is based on experimental measurements that describe the relationship between hemolysis, shear stress, and exposure time. It is generally known that with increased shear stress and or exposure time an increase in hemolysis is seen. To use this relationship in computational models, the experimental data is fitted using a power law approach $HI = C * \tau^\beta * t^\alpha$ providing a functional relationship between these variables. Despite its utility, the Power Law method primarily allows for relative comparisons rather than absolute hemolysis values. Numerous studies have attempted to refine this approach by investigating blood from different species [11–13] employing various apparatuses [13–16] or employing different numerical techniques [17–20]. However, no universal parameter set for the Power Law method has been established. A comprehensive overview of numerical hemolysis models is presented by reviews [21, 22].

A critical limitation of Power Law models is their inability to account for inherent variance in experimental data, as the fitting process results in a loss of this variance, leading to a representation that only captures an average scenario of hemolysis. This raises concerns about the robustness of the fitting approach used to determine the optimal model parameters (C, α, and β), as it should accurately mirror the original hemolysis signal of the underlying experiment. Notable attempts to address this robustness include Mohammadi et al.'s [23] adjustment of model parameters to calculate probability distributions of the model parameters and Craven et al.'s [24] device-specific hemolysis model, that relied on a combination of computational models and experimental device hemolysis data. Both studies highlighted a notable sensitivity of resulting hemolysis to changes in model parameters, underlining the importance of finding the correct parameters. However, both approaches lacked consideration of experimental variance in their models.

To overcome these limitations and meet the increased importance of reliable and robust hemolysis models incorporating UQ, this study aims to develop a universal method for analyzing all experimental hemolysis data sets, incorporating the variance of underlying experiments. By accounting for experimental variance, we can enhance the reliability and robustness of hemolysis predictions, ensuring that future regulatory submissions for blood-contacting medical devices are based on more robust and accurate models.

## 2. Materials and Methods

Figure 1 provides a visual summary of the analyses conducted and methods employed in this study. The study uses experimental data describing the relationship between Hemolysis Index (HI), exposure time (t), and shear stress ($\tau$) with a repetition number of n=3 at each operating point, utilizing a custom-built Couette-type shearing device. The exposure time ranged from 0.039s to 1.48s, and the shear stress ranged from 50 to 320 Pa, using ovine blood as the test species [25]. It compares the deterministic approach to determining hemolysis in a device with a novel probabilistic approach. At first the deterministic approach is analyzed in detail, revealing the relationship between experimental observations and model predictions. Then the probabilistic approach including the Bayesian parameter estimation method Markov Chain Monte Carlo (MCMC) and the reduced order model technique non-intrusive polynomial chaos expansion (NIPCE) is shown. In the end, the deterministic and probabilistic hemolysis models are compared using data from the FDA benchmark pump [26]. For easy reproducibility of the presented probabilistic approach, a python script to create the MCMC model as well as synthetic data to test the model can be found at https://zenodo.org/records/12820264.

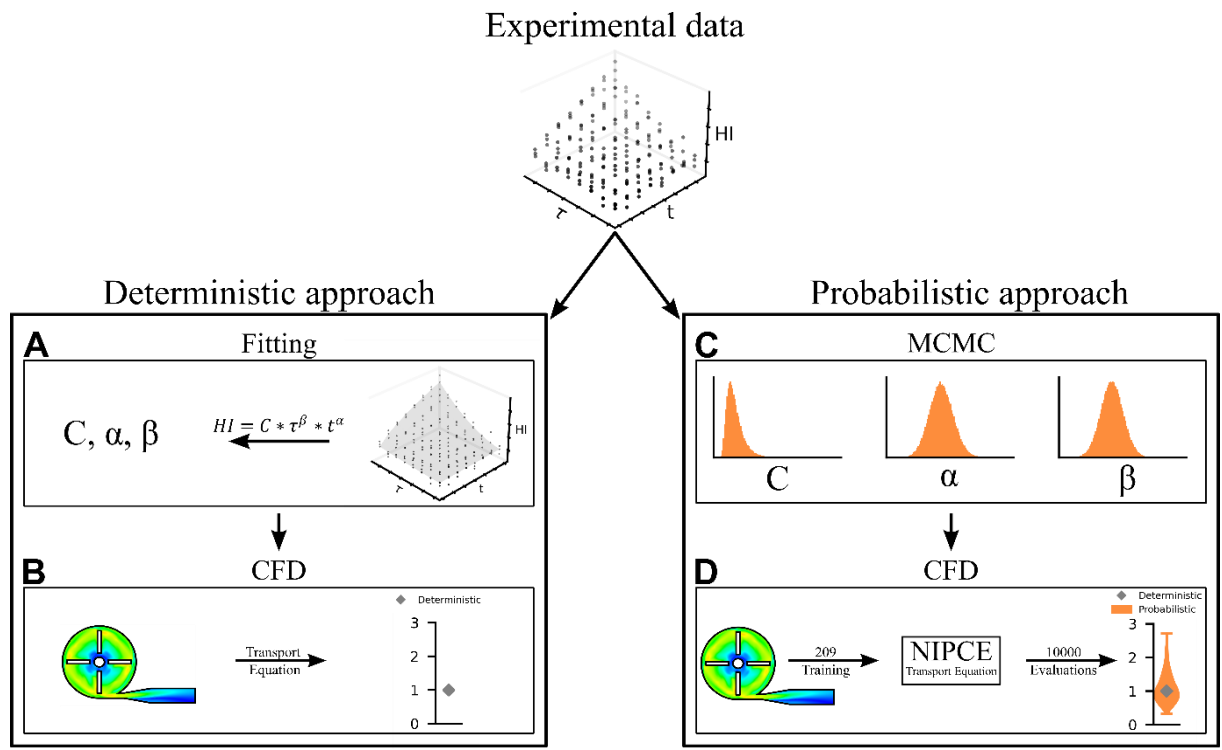

*Figure 1: Graphical overview of deterministic and novel probabilistic approach. Starting from experimental data A and B show the traditional deterministic approach with experimental data fitting and CFD solution steps respectively. In C and D the probabilistic approach is detailed with MCMC and CFD solution including NIPCE calculation, respectively.*

**Deterministic Approach**

The deterministic approach begins with fitting a Power Law model to the experimental data using Equation 1. For this, various fitting algorithms can be employed to achieve an optimal solution based on all experimental data, described by the optimal parameters C, α, and β of equation 1.

$$HI = C * \tau^\beta * t^\alpha \tag{1}$$

Equation 1 is then transformed into a transport equation and solved using Computational Fluid Dynamics (CFD) to obtain a deterministic hemolysis value (e.g., MIH) for a device under given boundary conditions.

For all results presented using the deterministic approach, the original model parameters determined by the multivariate regression method from Zhang et al. [25] were used (C = 1.228×10−5, α = 0.6606, β = 1.9918).

Regardless of the fitting method, the sum of squared errors (SSE) between the experimental observations and model predictions can be calculated:

$$SEE = \sum_{i=1}^{n}(y_i - \hat{y}_i)^2 \tag{2}$$

Here, $y_i$ represents the experimental observation of HI, and $\hat{y}_i$ denotes the model prediction of equation 1 with specific Parameters C, α, and β. The variable n describes the number of observed experimental values. To explore the possible parameter space and identify potential multiple optimal solutions, an initial global grid search with broader parameter ranges was conducted to determine appropriate ranges for a more detailed analysis. Subsequently, a refined grid search was performed over the narrowed values for C (1×10−5 to 1×10−3), α (0.1 to 1), and β (1.2 to 2.5), with 1000 values for α and β and 12 for C, resulting in 12,000,000 possible combinations that were analyzed.

**Probabilistic Approach**

The probabilistic approach uses the same experimental data as the deterministic approach, employing the MCMC method to determine the posterior distributions of the model parameters in Equation 1 with the highest likelihood of observing the provided experimental data. In contrast to optimal parameter values of the deterministic approach, which minimize one objective function and deliver one parameter set, this has several advantages. At first, Bayesian parameter estimation provide a global overview of possible parameter values and are not limited to local minima. Secondly, posterior parameter distributions can be updated given new experimental data or further information. At last, uncertainty quantification can be performed through sampling from the parameter distribution and running forward simulations.

The samples drawn from the posterior distributions of C, α, and β are transformed into hemolysis values using CFD, similar to the traditional approach. To achieve a meaningful posterior distribution of hemolysis values, this sampling process must be conducted very frequently. A reduced-order model of the CFD simulation, created using NIPCE, substantially accelerates the prediction of hemolysis based on the model parameters.

**MCMC Implementation**

MCMC methods in Bayesian statistics estimate posterior distributions of model parameters given observed data. This study used the Python package PyMC [27] to implement MCMC. Experimental data from Zhang et al. [25] and Equation 1 served as the basis for sampling posterior distributions. An additional parameter σ was introduced to capture the variability in the experimental data that cannot be explained by the model parameters C, α, and β. The initial distributions for C, α, β and $\sigma$ were assumed to be Uniform (0-1) for C and Normal (mean=0, std=1) for the other parameters.

The likelihood of observing the hemolysis index (HI) was modeled by a Student's t-distribution:

$$HI \sim t\left(\mu = \hat{y}_i, \sigma = \sigma, \nu = \frac{1}{30}\right) \qquad (3)$$

where μ represents the central value around which the data is distributed (equation 1), σ the spread of the data around the central value (measurement noise) and $\nu$ the heaviness of the tails (degrees of freedom) of the distribution following an exponential distribution with a rate parameter of 1/30. The Student's t-distribution was chosen for its robustness to outliers, effectiveness with small sample sizes, and ability to handle uncertainty in variance estimates. During MCMC sampling, $\sigma$ is sampled alongside C, α and β. The sampler explores the joint posterior distribution of all parameters, accounting for the interdependencies between them. This simultaneous sampling ensures that the uncertainty in $\sigma$ is properly propagated to the estimates of C, α and β. This means that $\sigma$ directly affects the probability of observing the data given the model parameters and that the posterior distributions of C, α and β incorporate the uncertainty in $\sigma$. Supporting information on this statement can be found in Figure SI 2.

Using the No-U-Turn Sampler (NUTS) algorithm [28], four chains with 50,000 samples each were sampled after a burn-in period of 1000 samples and with a target acceptance rate of 0.95. The resulting 200,000 samples per MCMC run provided the basis for the posterior distributions. Chain convergence was assessed using trace plots and Gelman-Rubin Diagnostic, with an example trace plot provided in the Supplementary Information (Figure SI 1).

Models with a constant C parameter were also computed. For different values of C, the distributions of $\alpha$, $\beta$, and $\sigma$ took on different forms. Bayesian optimization identified the C value that yielded the smallest median value of the $\sigma$ distribution. For this C was varied within the bounds of $1\times10^{-6}$ to $1\times10^{-3}$, and the optimal value of C=3.515e-5 was found within 100 iterations using the *gp_minimize* function of the Python package *scikit-optimize [29]* (Figure SI 5).

**CFD Model**

To compare the probabilistic and deterministic methods, the study utilized the FDA Round Robin benchmark blood pump [26]. This blood pump has been extensively investigated for hemolysis in a multicenter study [30, 31], making it an ideal basis for comparing hemolysis predictions from the models with experimental and simulation data. The validation of the Computational Fluid Dynamics (CFD) model was previously conducted by our research group, as detailed by Gross-Hardt et al. [32]. For comprehensive information regarding the mesh and simulation parameters, please refer to their publication.

To summarize, the geometry was meshed using the ANSYS 2021 R1 Meshing tool (ANSYS Inc., Canonsburg, USA), employing unstructured tetrahedral elements with prism layers. This meshing approach resulted in a total element count of 9.3 million cells, corresponding to the "medium" mesh configuration described in the earlier study [32] . The CFD simulations were performed using the ANSYS CFX solver, with the k-$\omega$ shear stress transport model selected for turbulence modeling. Unlike Gross-Hardt et al. [32], steady-state simulations were conducted to reduce computational effort for generating training data. This approach was validated against experimental data [30], demonstrating that the simplification to steady-state simulations is justified due to the minimal differences observed when compared to unsteady simulations and the validation dataset. For hemolysis calculations the Eulerian method according to Garon and Farinas [33] with the model parameters of Zhang et al. [25] was used.

**Reduced-Order Model**

To expedite the propagation of experimental uncertainty from the MCMC posterior distributions to hemolysis values, Non-Intrusive Polynomial Chaos Expansion (NIPCE) was applied. The MCMC posterior distributions were fitted using log-normal distributions truncated at the lower 1% and upper of the 99% intervals to avoid extreme values in the subsequent sampling process. These distributions were then combined into a multivariate distribution. Next, 200 samples were drawn from this multivariate distribution using Latin Hypercube Sampling. These samples, along with combinations of the maximum and minimum values of the individual distributions, were used as training data for the reduced-order model, resulting in either 209 or 204 training points depending on whether C, α, and β, or just α and β were used. Hemolysis was determined for these parameter combinations using the CFD approach described above. The resulting dataset of model parameters and corresponding hemolysis values was used to train the NIPCE model with a polynomial order of 4. This model can transform a large number of MCMC model parameter samples into a distribution of hemolysis values within seconds. More information about this reduced order model technique in the context of blood pumps can be found in [34] and more information as well as validation of the training process is provided in the Supplementary Information (Figure SI 3 and SI 4). Generating the training data points using CFD on a standard laptop took approximately 60 seconds, and NIPCE training with these samples took 10 seconds. Hemolysis with uncertainty quantification was determined through 10000 forward simulations of NIPCE for each simulation condition taking less than a second for all 6 conditions.

**Model comparison for FDA pump setup**

To compare the deterministic and probabilistic models, experimental hemolysis data of Conditions 1-6 from Ponnaluri et al. [31] was used. RIH values were extracted from Figure 11 c) using a graph digitization tool. According to Equation 4 in Ponnaluri et al. [31], the RIH values correspond to the modified hemolysis index (MIH) values, normalized to Condition 5 of the study. For the probabilistic approach, this normalization involved dividing by the median of the posterior MIH distribution of Condition 5.

## 3. Results

Figure 2A shows discrete contour slices of the Sum of Squared Errors (SSE) in a three-dimensional plot for the parameter space α (0.1 to 1), and β (1.2 to 2.5), with 12 fixed, equidistant C values in the interval (1×10−5 to 1×10−3). The size of the low SSE regions decreases from high to low C values. To maintain a similar SSE at a given C level, β can only vary slightly (±0.02), while α can vary more (±0.3). This indicates that the optimal fit for HI is more sensitive to β than to α. Notably, despite different C levels, several regions exhibit similarly low SSE values between 3.5 and 3.6.

This is further highlighted in Figure 2B, displaying the optimal SSE value for each C level in solid black, along with the modified index of hemolysis (MIH) of the FDA pump setup (Condition 5) for each optimal SSE value in dashed black. Despite similar SSE values, the absolute value of hemolysis varies by approximately 30%. This suggests that the SSE function, derived from Equation 1 and the experimental data, has multiple local minima across different parameter ranges which can significantly impact the absolute hemolysis level.

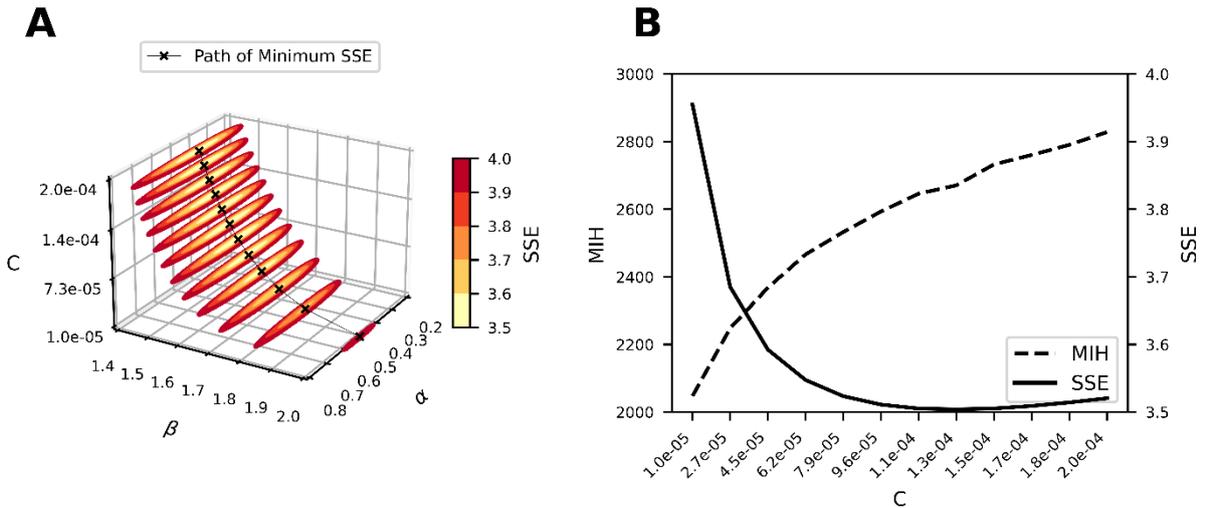

*Figure 2: Three-dimensional contour plots depicting the Sum of Squared Errors (SSE) of experimental data against the power law model across the parameters C, α, and β. The black line with cross symbols indicates the minimum SSE value for each level of C (panel A). In panel B the minimum SSE value for each C level is plotted in solid black and the corresponding Modified Index of hemolysis (MIH) for Condition5 of the FDA blood pump is shown in dashed black.*

Figure 3A presents the results of the MCMC method in the form of a corner plot. It displays the individual distributions of the parameters C, α, and β, as well as their relationships in a combination of scatter and contour plots. The contour levels of the 95%, 50%, and 5% confidence intervals are depicted in yellow, orange and red inside the scatter plots, respectively.

It is clearly visible that the parameters C and β are correlated. This implies that no unique solution is possible, indicating redundancy in the model. An increase in one parameter can be compensated by the decrease in the other one. This observation indicates that defining C with a constant value and fitting the other two remaining parameters is a better strategy than optimizing all three parameters.

In Figure 3B, the corner plot with a constant C = 1.228e-5, as used in the original Zhang et al. [25] model, is shown. It can be observed that through MCMC sampling, the median of the α posterior distribution has settled at 0.632, and the median of the β posterior distribution at 1.985. Additionally, no correlation between the parameters is evident.

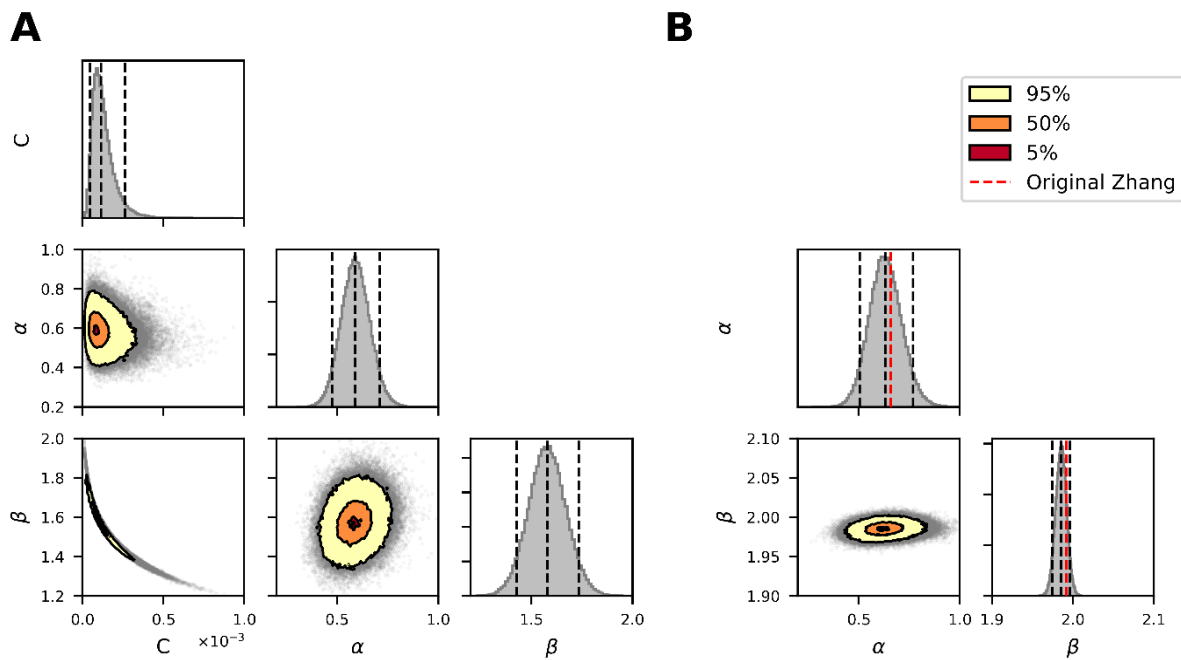

*Figure 3: Displays the corner plot of the MCMC method for the power law model with parameters C, α, and β in panel A, and for the model with fixed C=1.228e-5, with variable α and β in panel B. The histograms show the median, along with the 5% and 95% intervals, indicated by dashed lines. The scatter plots are overlaid with contour plots representing the 95%, 50%, and 5% confidence intervals.*

However, with a constant C model, it is not guaranteed that the experimental data can be optimally reproduced. To determine the constant C value that best fits the experimental data, a Bayesian optimization was conducted on the parameter $\sigma$. Figure 4 shows in panels A-C the various distributions of α, β, and $\sigma$ with different C values ranging from 1e-8 to 1e-1. It is evident that as the C value increases, the median value of the β distribution decreases, while the width remains relatively constant. For $\sigma$, which describes the differences between the observed values and the values predicted by the model, it is observed that there must be a minimal value between C=1e-8 and C=1e-1. The minimization of the median of the $\sigma$ distribution through Bayesian optimization (Figure SI 4) leads to an optimal C value of 3.515e-5.

The parameter distributions of this optimal model, providing the best fit to the data and minimizing the residual error between the observed and predicted hemolysis values, are shown in Figure 4 D-F. Alongside the histogram created from the MCMC samples, the 5% and 95% interval boundaries as well as the median values are indicated with dashed lines. With the given shape and scale parameters of the log-normal distribution, as well as the constant location parameter set to 0, the depicted distributions of the optimal model can be accurately reproduced.

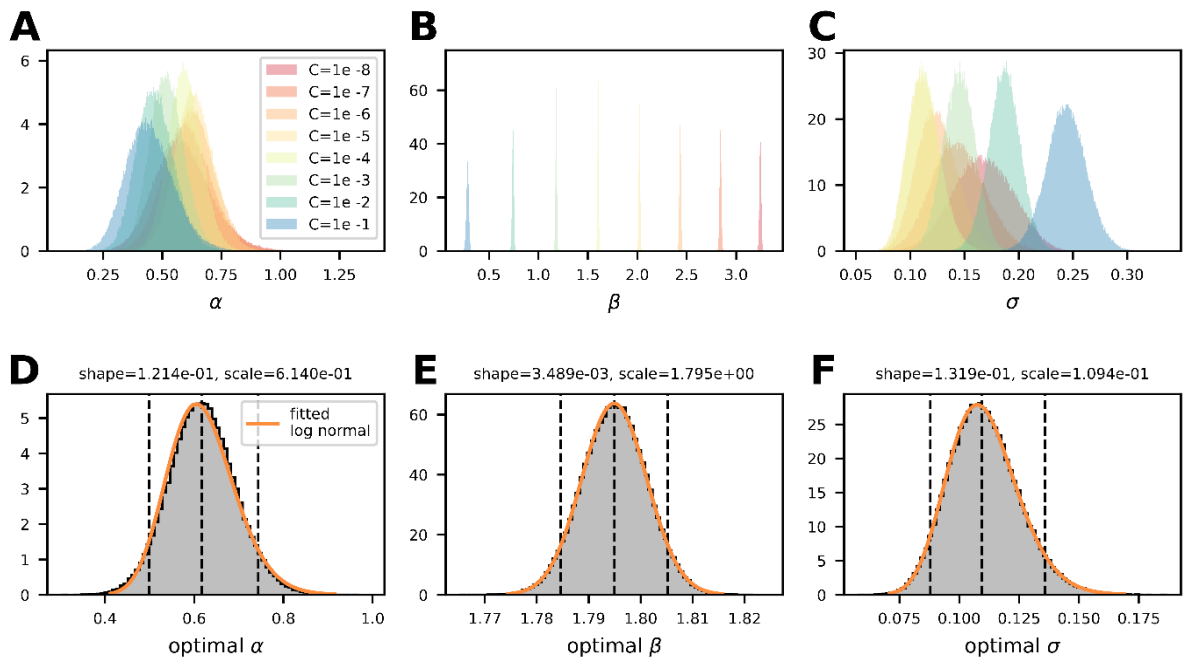

*Figure 4: Panels A-C illustrate the influence of different constant C levels, ranging from 1 to 1e-8, on the parameter distributions of α, β, and σ, respectively. Panels D-F depict the parameter distributions for the optimal C=3.515e-5 for α, β, and σ. The dashed lines indicate the 5%, 95%, and median values of the individual distributions.*

This optimal model is compared to the traditional deterministic approach and experimental data from the FDA round robin study [26] in Figure 5. The relative hemolysis index (RIH) with respect to Condition 5 is plotted for boundary conditions 1-6. The comparison between the mean values of the experimental data from Ponnaluri et al. [31] and the deterministic data of Zhang et al. [25] (black diamonds) shows that all deterministically calculated RIH values lie within the experimental measurements (mean ± SD). The new probabilistic method, represented by violin plots, shows a similar pattern regarding the median values of the distribution. In addition to the median values, the distribution of possible hemolysis values can be observed through the propagated uncertainty of the underlying experimental data. It is also evident that Conditions 1 and 4, which correspond to a rotational speed of 2500 rpm, have a smaller distribution width compared to the other conditions with a rotational speed of 3500 rpm.

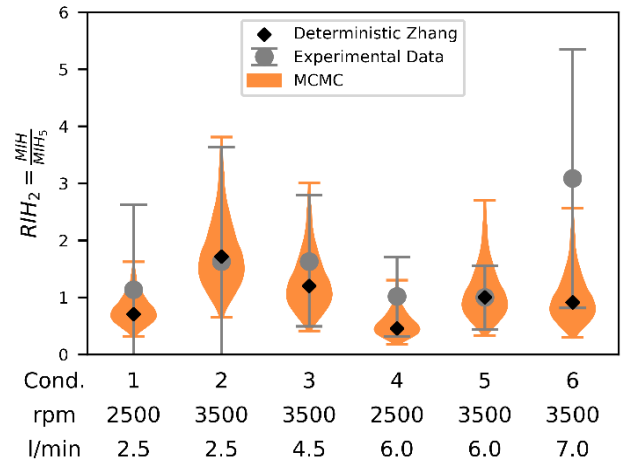

*Figure 5: This figure compares the deterministic and probabilistic approaches with the experimental data from the FDA round robin study under six different operating conditions [31]. The deterministic model [25] is represented by black diamonds, the probabilistic model by orange violin plots, and the experimental data's mean ± SD is shown with grey error bars.*

## 4. Discussion

The aim of this study was to develop a universally applicable method that transfers the uncertainties from the underlying experimental data of hemolysis models into the prediction of hemolysis outcomes. Through a detailed analysis of the traditional Power Law approach, as well as the integration of reduced-order models and Bayesian statistical methods, three main findings were identified:

1. For the traditional C, α, β Power Law, due to the correlation between C and β, an optimal, unique fitting of model parameters cannot be found.
2. Even with a constant C, multiple possibilities for α and β exist to achieve an optimal fit due to the variance in experimental data.
3. Considering the variance from the underlying experiments enhances the robustness of hemolysis model predictions and simplifies the comparison between experimental and simulated data.

As illustrated in Figures 2 and 3, there is a clear correlation between the parameters C and β of the Power Law equation (Equation 1). This relationship results in an inability to find a globally unique solution, indicating redundancy in the model (one parameter increases while the other decreases to compensate). Craven et al. [24] have previously found that there are infinitely many different parameter combinations for a single operating point that lead to the same hemolysis value. This study extends that analysis, demonstrating that the proposed Power Law function is generally unsuitable for finding a unique optimal solution to the fitting process given experimental data [25]. Consequently, different parameter combinations can yield similarly good fits to the experimental data but result in significantly different absolute hemolysis values due to their parameter differences (Figure 2).

To the best of the authors' knowledge, this is the first study to incorporate the variance in the underlying experiments of hemolysis models into the fitting process of model parameters. Despite fixing the model parameter C and thus reducing the possible parameter space, finding the optimal fit for parameters remains challenging. As shown in various slices from Figure 2, due to experimental variance, a large number of different α and β parameters can achieve an optimal fit with similar SSE values.

Utilizing the MCMC method from Bayesian statistics, we have developed a hemolysis model that integrates experimental variance by employing probability distributions for the model parameters. To address the problem of correlation between C and β, C was fixed as a constant, and an optimal C was found using Bayesian optimization to maximize the agreement between the model and experimental data, thereby minimizing $\sigma$. In this context, minimizing $\sigma$ refers to reducing the variability of the residuals (the differences between observed and predicted values). This reduction in residual variability enhances the precision of our model estimates, ensuring that the uncertainty in the model parameters $\alpha$ and $\beta$ is as low as possible. Consequently, the probability distributions of $\alpha$ and $\beta$ more accurately reflect the variance of the underlying experimental data, leading to a more reliable model. This approach differs significantly from previous literature [23], which randomly samples probability distributions for $\alpha$, $\beta$, and C from parameter intervals derived from literature values of multiple models.

Our approach strengthens the robustness of relative hemolysis predictions by considering potential variance factors such as inter- or intra-variability of blood samples from the underlying experiments, which can also be sources of variance in the experimental investigation of MCS devices. A test of the MCMC method using artificially generated experimental data, which strictly follow the surface defined by the original model parameters of Zhang et al. [25], demonstrates that distributions of $\alpha$ and $\beta$ observed in this study are solely due to the variance in the underlying data(Figure SI 2). The probabilistic representation of hemolysis values as a probability distribution enhances comparability to experimental results, which, as seen from the FDA pump data [31], also show notable variance. Furthermore, the already known strong sensitivity to different values of model parameters [23, 24, 35] on the predicted hemolysis values is directly visible through the distribution representation of our model, allowing the model results to be interpreted accordingly.

In comparison, experimental FDA pump data may be influenced by numerous other factors that contribute to the measured variance in hemolysis values, which cannot all be covered in a numerical model. It is important to note that the raw data from Zhang et al. [25] examined bovine blood, while the FDA pump experiments [31] investigated porcine blood. Consequently, the comparison to the FDA pump data serves only as an illustrative example of how the probabilistic hemolysis model simplifies the comparison to experimental data. Nonetheless, despite the different donors, a relative comparison between simulation and experiment is still possible.

However, the problem of accurately predicting absolute hemolysis values is not solved by the probabilistic model. Absolute hemolysis values are within the same order of magnitude as those predicted by classical power law models. This issue does not appear to be due to the variance in the underlying experimental data but, as discussed in the literature, seems to result from more fundamental modeling issues.

This study highlights a fundamental mathematical limitation of the fitting process in the classic power law approach and successfully demonstrates how the inherent uncertainty in hemolysis experiments can be captured and implemented into numerical blood damage models. It further shows that incorporating fitting parameter variability through MCMC substantially enhances the robustness of hemolysis model predictions. The current gold standard of relative comparisons is strengthened by incorporating the variance of the underlying experiments, providing a stronger foundation for comparing simulated hemolysis outcomes with in-vivo experiments. The developed method can easily incorporate further experimental datasets encompassing various stress types, donor species, and a higher number of repetitions. Future work could include more data points from different experiments with varied shear conditions and exposure times. Due to the nature of the probabilistic framework, this can be easily done by updating the posterior distributions to the new data. This approach has the potential to result in a comprehensive model that incorporates all the experimental knowledge of the scientific community, setting a new standard of predictive accuracy in hemolysis modeling.

# References

1. Shapira Y, Vaturi M, Sagie A. Hemolysis associated with prosthetic heart valves: a review. Cardiol Rev. 2009;17:121–4. doi:10.1097/CRD.0b013e31819f1a83.
2. Shah P, Tantry US, Bliden KP, Gurbel PA. Bleeding and thrombosis associated with ventricular assist device therapy. J Heart Lung Transplant. 2017;36:1164–73. doi:10.1016/j.healun.2017.05.008.
3. Halaweish I, Cole A, Cooley E, Lynch WR, Haft JW. Roller and Centrifugal Pumps: A Retrospective Comparison of Bleeding Complications in Extracorporeal Membrane Oxygenation. ASAIO J. 2015;61:496–501. doi:10.1097/MAT.0000000000000243.
4. Sniderman J, Monagle P, Annich GM, MacLaren G. Hematologic concerns in extracorporeal membrane oxygenation. Res Pract Thromb Haemost. 2020;4:455–68. doi:10.1002/rth2.12346.
5. Omar HR, Mirsaeidi M, Socias S, Sprenker C, Caldeira C, Camporesi EM, Mangar D. Plasma Free Hemoglobin Is an Independent Predictor of Mortality among Patients on Extracorporeal Membrane Oxygenation Support. PLoS One. 2015;10:e0124034. doi:10.1371/journal.pone.0124034.
6. Nunez JI, Gosling AF, O'Gara B, Kennedy KF, Rycus P, Abrams D, et al. Bleeding and thrombotic events in adults supported with venovenous extracorporeal membrane oxygenation: an ELSO registry analysis. Intensive Care Med. 2022;48:213–24. doi:10.1007/s00134-021-06593-x.
7. Gacic M, Kaplarevic M, Filipovic N. Cost-effectiveness analysis of in silico clinical trials of vascular stents. In: 2021 IEEE 21st International Conference on Bioinformatics and Bioengineering (BIBE); 25.10.2021 - 27.10.2021; Kragujevac, Serbia: IEEE; 10252021. p. 1–5. doi:10.1109/BIBE52308.2021.9635321.
8. Badano A. In silico imaging clinical trials: cheaper, faster, better, safer, and more scalable. Trials. 2021;22:64. doi:10.1186/s13063-020-05002-w.
9. Musuamba FT, Skottheim Rusten I, Lesage R, Russo G, Bursi R, Emili L, et al. Scientific and regulatory evaluation of mechanistic in silico drug and disease models in drug development: Building model credibility. CPT Pharmacometrics Syst Pharmacol. 2021;10:804–25. doi:10.1002/psp4.12669.

10. U.S. Food and Drug Administration. Assessing the Credibility of Computational Modeling and Simulation in Medical Device Submissions - Guidance for Industry and Food and Drug Administration Staff.

11. M. Giersiepen, L.J. Wurzinger, R. Opitz, and H. Reul. Estimation of Shear Stress-related Blood Damage in Heart Valve Prostheses - in Vitro Comparison of 25 Aortic Valves.

12. Ding J, Niu S, Chen Z, Zhang T, Griffith BP, Wu ZJ. Shear-Induced Hemolysis: Species Differences. Artif Organs. 2015;39:795–802. doi:10.1111/aor.12459.

13. Heuser G OR. A Couette viscometer for short time shearing of blood. Biorheology;1980.

14. Fraser KH, Zhang T, Taskin ME, Griffith BP, Wu ZJ. A quantitative comparison of mechanical blood damage parameters in rotary ventricular assist devices: shear stress, exposure time and hemolysis index. J Biomech Eng. 2012;134:81002. doi:10.1115/1.4007092.

15. Mei X, Lu B, Wu P, Zhang L. In vitro study of red blood cell and VWF damage in mechanical circulatory support devices based on blood-shearing platform. Proc Inst Mech Eng H. 2022;236:860–6. doi:10.1177/09544119221088420.

16. Froese V, Goubergrits L, Kertzscher U, Lommel M. Experimental validation of the power law hemolysis model using a Couette shearing device. Artif Organs. 2024;48:495–503. doi:10.1111/aor.14702.

17. Dirkes N, Key F, Behr M. Eulerian formulation of the tensor-based morphology equations for strain-based blood damage modeling. Computer Methods in Applied Mechanics and Engineering. 2024;426:116979. doi:10.1016/j.cma.2024.116979.

18. Gesenhues L, Pauli L, Behr M. Strain-based blood damage estimation for computational design of ventricular assist devices. Int J Artif Organs. 2016;39:166–70. doi:10.5301/ijao.5000484.

19. Chen Y, Sharp MK. A strain-based flow-induced hemolysis prediction model calibrated by in vitro erythrocyte deformation measurements. Artif Organs. 2011;35:145–56. doi:10.1111/j.1525-1594.2010.01050.x.

20. Arwatz G, Smits AJ. A viscoelastic model of shear-induced hemolysis in laminar flow. Biorheology. 2013;50:45–55. doi:10.3233/BIR-130626.

21. Faghih MM, Sharp MK. Modeling and prediction of flow-induced hemolysis: a review. Biomech Model Mechanobiol. 2019;18:845–81. doi:10.1007/s10237-019-01137-1.

22. Yu H, Engel S, Janiga G, Thévenin D. A Review of Hemolysis Prediction Models for Computational Fluid Dynamics. Artif Organs. 2017;41:603–21. doi:10.1111/aor.12871.

23. Mohammadi R, Karimi MS, Raisee M, Sharbatdar M. Probabilistic CFD analysis on the flow field and performance of the FDA centrifugal blood pump. Applied Mathematical Modelling. 2022;109:555–77. doi:10.1016/j.apm.2022.05.016.
24. Craven BA, Aycock KI, Herbertson LH, Malinauskas RA. A CFD-based Kriging surrogate modeling approach for predicting device-specific hemolysis power law coefficients in blood-contacting medical devices. Biomech Model Mechanobiol. 2019;18:1005–30. doi:10.1007/s10237-019-01126-4.
25. Zhang T, Taskin ME, Fang H-B, Pampori A, Jarvik R, Griffith BP, Wu ZJ. Study of flow-induced hemolysis using novel Couette-type blood-shearing devices. Artif Organs. 2011;35:1180–6. doi:10.1111/j.1525-1594.2011.01243.x.
26. Hariharan, P., Giarra, M., Reddy, V., Day, S. W., Manning, K. B., Deutsch, S., Stewart, S. F. C., Myers, M. R., Berman, M. R., Burgreen, G. W., Paterson, E. G., and Malinauskas, R. A. Multilaboratory Particle Image Velocimetry Analysis of the FDA Benchmark Nozzle Model to Support Validation of Computational Fluid Dynamics Simulations. ASME. J Biomech Eng;2011:133(4): 041002.
27. Abril-Pla O, Andreani V, Carroll C, Dong L, Fonnesbeck CJ, Kochurov M, et al. PyMC: a modern, and comprehensive probabilistic programming framework in Python. PeerJ Comput Sci. 2023;9:e1516. doi:10.7717/peerj-cs.1516.
28. Hoffman MD., Gelman A. The No-U-Turn Sampler: Adaptively Setting Path Lengths in Hamiltonian Monte Carlo. Journal of Machine Learning Research. 2014:1593–623.
29. Tim Head,MechCoder,Gilles Louppe,Iaroslav Shcherbatyi,fcharras,Zé Vinícius,cmmalone,Christopher Schröder,nel215,Nuno Campos,Todd Young,Stefano Cereda,Thomas Fan,rene-rex,Kejia (KJ) Shi,Justus Schwabedal,carlosdanielcsantos,Hvass-Labs,Mikhail Pak,SoManyUsernamesTaken,Fred Callaway,Loïc Estève,Lilian Besson,Mehdi Cherti,Karlson Pfannschmidt,Fabian Linzberger,Christophe Cauet,Anna Gut,Andreas Mueller,Alexander Fabisch. scikit-optimize/scikit-optimize: v0.5.2. Zenodo. 2018.
30. Malinauskas RA, Hariharan P, Day SW, Herbertson LH, Buesen M, Steinseifer U, et al. FDA Benchmark Medical Device Flow Models for CFD Validation. ASAIO J. 2017;63:150–60. doi:10.1097/MAT.0000000000000499.
31. Ponnaluri SV, Hariharan P, Herbertson LH, Manning KB, Malinauskas RA, Craven BA. Results of the Interlaboratory Computational Fluid Dynamics Study of the FDA Benchmark Blood Pump. Ann Biomed Eng. 2023;51:253–69. doi:10.1007/s10439-022-03105-w.

32. Gross-Hardt SH, Sonntag SJ, Boehning F, Steinseifer U, Schmitz-Rode T, Kaufmann TAS. Crucial Aspects for Using Computational Fluid Dynamics as a Predictive Evaluation Tool for Blood Pumps. ASAIO J. 2019;65:864–73. doi:10.1097/MAT.0000000000001023.
33. Garon A FMI. Fast three-dimensional numerical hemolysis approximation. Artif Organs. 2004.
34. Blum C, Steinseifer U, Neidlin M. Systematic analysis of non-intrusive polynomial chaos expansion to determine rotary blood pump performance over the entire operating range. Comput Biol Med. 2024;168:107772. doi:10.1016/j.compbiomed.2023.107772.
35. Mantegazza A, Tobin N, Manning KB, Craven BA. Examining the universality of the hemolysis power law model from simulations of the FDA nozzle using calibrated model coefficients. Biomech Model Mechanobiol. 2023;22:433–51. doi:10.1007/s10237-022-01655-5.